\documentstyle[12pt,epsf,world_sci]{article}
\newcommand{\geapp}{{\rm\raisebox{-0.7ex}{$\stackrel{\textstyle >}{\sim}$}}}
\begin{document}
\begin{flushright}
\vspace*{-0.6in}
                                                     FERMILAB-CONF-94/243-E \\
                                                       Submitted to DPF '94 \\
\end{flushright}
\title{NEW SHOWER MAXIMUM TRIGGER FOR ELECTRONS AND PHOTONS
AT CDF\thanks{Presented
        by David Gerdes, representing the CDF Collaboration.} }
\author{D.~AMIDEI, K.~BURKETT, D.~GERDES, C.~MIAO, D.~WOLINSKI \\
        {\em Randall Laboratory of Physics, University of Michigan, \\
             500 E. University Avenue, Ann Arbor, MI 48109} \\
     \vspace{0.3cm}
     and \\
     \vspace{0.3cm}
        K.~BYRUM, J.~DAWSON, L.~NODULMAN, A.~B.~WICKLUND \\
        {\em Argonne National Laboratory, Argonne, IL 60439} \\ }

\maketitle
\setlength{\baselineskip}{2.6ex}

\begin{center}
\parbox{13.0cm}
{\begin{center} ABSTRACT \end{center}
{\small \hspace*{0.3cm}
For the 1994 Tevatron collider run, CDF has upgraded the
electron and photon trigger hardware to make use of shower position and size
information from the central shower maximum detector. For electrons,
the upgrade has resulted in a 50\% reduction in backgrounds while
retaining approximately 90\% of the signal. The new trigger also
eliminates the background to photon triggers from single-phototube discharge.}}
\end{center}

Inclusive electron triggers have provided CDF with a rich stream of
data. Studies of the $W$ and $Z$ bosons, as well as the search for and
study of the top quark, are carried out in part using electron triggers
with a threshold of typically 12-16 GeV. Topics in $b$ physics, including
production properties, the identification of exclusive decays,
lifetime studies, and in
particular the search for rare processes, demand a lower trigger threshold.
In addition, the low-threshold electron trigger provides important
calibration samples.
In the 1988-89 and 1992-93 Tevatron collider runs, this threshold
varied between 6 and 9~GeV, and the lower-threshold trigger was prescaled.
With the long-anticipated, and now realized\cite{lum}, high-luminosity
conditions of the 1994-95 collider run, a means for reducing the cross
section of this trigger without raising the threshold was necessary in
order to keep the rate of
accepted events from becoming unmanageably high. We have solved this
problem by making shower position and size information from the
central strip chambers (CES) available for use in the trigger decision.

The CDF detector and trigger system have been described in detail
elsewhere.\cite{cdf_det,cdf_trig} The detector components of interest
for this discussion are the central tracking chamber (CTC), the central
electromagnetic and hadronic calorimeters (CEM and CHA), and the CES.
The CTC, which is located inside a 1.4-T solenoidal magnetic field,
is a cylindrical drift chamber with 84 layers, grouped
into five axial and four stereo superlayers. Fast timing information from
the axial layers is used by a hardware track finder, the Central Fast
Tracker (CFT), which has
a transverse momentum resolution of $\delta P_T/P_T = 3.5\%\times P_T$.
The CEM and CHA, located outside the solenoid,
cover the pseudorapidity region $|\eta|<1.1$, and are organized into
projective towers in $\eta$-$\phi$ space of size 0.1$\times 15^{\circ}$,
where $\phi$ is the azimuthal angle. Fast analog outputs on these calorimeters
make the energy in ``trigger towers" of size 0.2$\times 15^{\circ}$ available
for immediate use by the trigger. The CES, located inside the CEM
near EM shower maximum at
a depth of six radiation lengths, provides
shower position and amplitude information in both the $r$-$\phi$ and
$z$ views. Until the present upgrade, this information was not available
to the hardware triggers.

The hardware electron trigger consists of two levels. Level-1
operates without deadtime in the 3.5~$\mu$sec window between beam crossings,
and requires at least 8~GeV of transverse energy, $E_T$, in a CEM trigger
tower. This requirement also serves as the photon trigger at Level-1.
At level-2, a hardware cluster-finder identifies clusters with
at least 87.5\% electromagnetic energy, and a CFT track is sought that
matches this cluster in momentum and azimuth. (The photon trigger
works similarly, but with an independent threshold and no track-match
requirement.) Events that satisfy these requirements are accepted as
electron candidates, and are passed to the third level of the trigger,
which is a software trigger that uses the full detector information.
The rate of events into Level-3 is bandwidth-limited, hence the desirability
of reducing the rate out of Level-2.

Fewer than 10\% of events that pass the Level-2 electron trigger
actually contain good primary electrons. The remaining events consist of
conversions, hadronic showers that fluctuate into predominantly
electromagnetic energy, and
$\pi^0$-$\pi^{\pm}$ overlaps (where the neutral pion
provides the electromagnetic shower and the charged pion provides the
track). While conversions constitute a valuable control sample,
the remaining backgrounds can be significantly
reduced by making use of shower position information from the CES.
This is because the default trigger can make only a 15$^{\circ}$ match
between the track and the EM cluster, much too loose to reject overlaps.
In addition, the default trigger does not know the depth in the CEM
at which energy was deposited, so it cannot reject
early hadronic showers that take place in the back portion of the CEM,
after EM shower maximum. Similarly, the background to photon triggers
from single-phototube discharges, which average approximately 1~Hz out of the
overall 25-50~Hz allotted for all level-2 accepts, can be rejected by
requiring energy in the CES.

We have therefore constructed new front-end readout
cards\cite{rabbit} (known as XCES boards) for the CES
that make the shower position information available in Level-2, for
CES clusters above a selectable threshold. The XCES boards perform
sums of the energy on groups of four adjacent CES wires, corresponding to
a 2$^{\circ}$ $\phi$ segment, and compare them to a threshold
(typically $\approx 4$~GeV)
supplied by an on-board adjustable DAC.
The resulting on/off bits,
eight bits for each of the 24 15$^{\circ}$ wedges in each half of
the detector, for a total of $8\cdot 24\cdot 2 = 384$~bits, are
latched by additional new trigger hardware (the CERES board) following a
level-1 accept. The turn-on curve for the XCES bits, as a function of
the energy deposited on the four wires, is shown in Figure~1.
\begin{figure}
\begin{minipage}[t]{2.9in}
 \epsfxsize=2.6in
 \epsfbox[0 162 500 657]{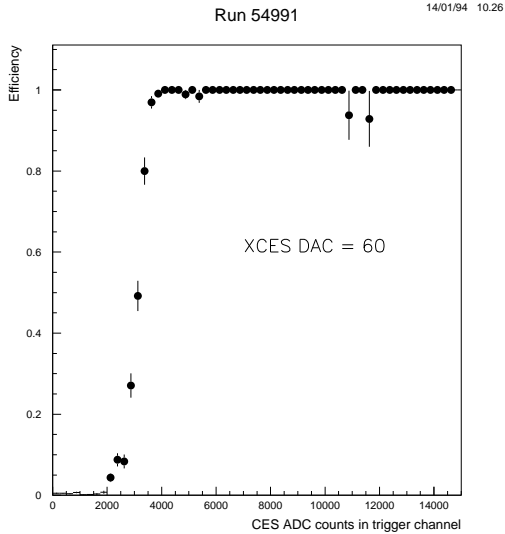}
 \label{turnon}
 \caption{The turn-on efficiency for XCES bits as a function of the
          CES energy deposition in a 4-wire strip, expressed here in
          ADC counts. 1 GeV $\approx$ 800 counts. The DAC threshold
          that determines the turn-on point is adjustable and is
          shown here at its nominal value.}
 \end{minipage}
\end{figure} \ \
\begin{figure}
\hspace*{3.1in}
\begin{minipage}[t]{2.9in}
\vspace*{-4.23in}
 \epsfxsize=2.6in
 \epsfbox[0 162 500 657]{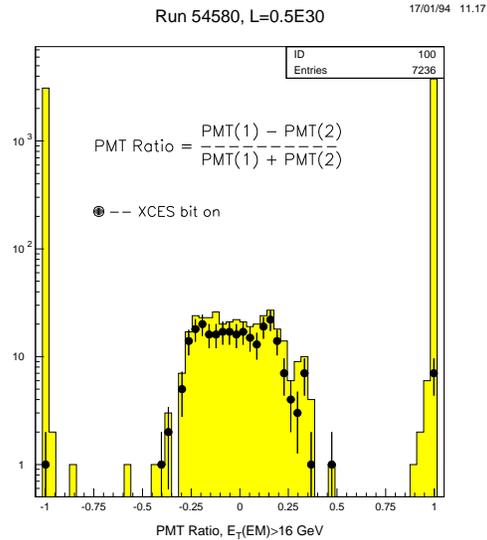}
 \label{single_pmt}
 \caption{The phototube ratio $R$, defined as the difference between the
          two phototube energies in a tower divided by their sum.
          Single-phototube discharges show up at $\pm$1, and dominate the
          physics rate in this low-luminosity run. The points show
          $R$ for towers with an OR-bit turned on.}
\end{minipage}
\end{figure}

The CERES board, a double-width, surface-mounted Fastbus board, receives
the XCES bits along with track $\phi$ and signed $P_T$ information
from the CFT. A large lookup table is used to identify tracks
that match to a CES cluster, and these tracks are flagged for use in
the electron trigger. The electron trigger then operates as before,
with the additional requirement that the track associated to the EM
cluster must have been flagged by the XCES/CERES system. The CERES
board also performs an OR of the eight XCES bits from each wedge.
The resulting 48 bits are used to reject the single-phototube
background to the photon trigger by requiring that the CEM tower that
gave the trigger also have the relevant ``OR-bit" set. Figure~2
shows the near-total rejection obtained by this requirement.

As implemented in the CDF trigger for the current collider run,
the XCES-based electron trigger requires an EM cluster with $E_T>8$~GeV,
associated to a CFT track with $P_T>7.5$ GeV that is required
to match to a CES cluster. At luminosities above $10^{31}$cm$^{-2}$s$^{-1}$
the trigger is prescaled by a factor of 2.
This trigger has two adjustable parameters:
the CES threshold at which XCES bits are set, and the road width used in
the CERES lookup table to define a track-cluster match. We have varied
these parameters in a series of studies in which the Level-3 trigger
was operated in tagging mode.

The cross section for the 8~GeV trigger is shown in Figure~3
as a function of the road size used in the lookup table.
The nominal CES threshold of 3500 ADC counts (see Figure~1) was used for
this study. For comparison, the cross section for
this trigger with no XCES requirement is also shown. Even with a very
wide road the rate is reduced by a factor of 1.4, indicating the effect
of requiring a CES cluster above threshold anywhere in the wedge.
For the nominal road size of 3~cm the trigger cross section is reduced
by more than a factor of two, from 840~nb to 400. The efficiency
for the nominal threshold and road sizes, measured using
a control sample of electrons from photon conversions,
is shown in Figure~4. The efficiency is $\approx 85\%$ at the trigger
threshold, and rises to 100\% for $E_T\geapp 13$~GeV.
We find little gain in efficiency for larger road sizes, and attribute the
increase in the cross section to $\pi^0$-$\pi^{\pm}$ overlaps.
\begin{figure}
\begin{minipage}[t]{2.9in}
 \epsfxsize=2.7in
 \epsfbox[0 162 500 657]{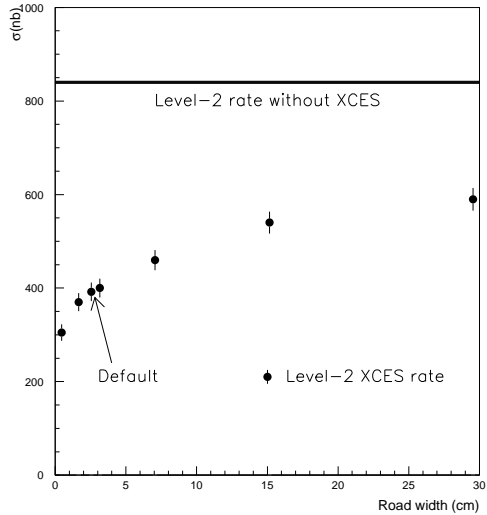}
 \label{rate_vs_road}
 \caption{Cross section for the 8~GeV electron trigger as a function
          of the road size used to match tracks to CES clusters.}
 \end{minipage}
\end{figure} \ \
\begin{figure}
\hspace*{3.1in}
\begin{minipage}[t]{2.9in}
\vspace*{-3.52in}
 \epsfxsize=2.7in
 \epsfbox[0 162 500 657]{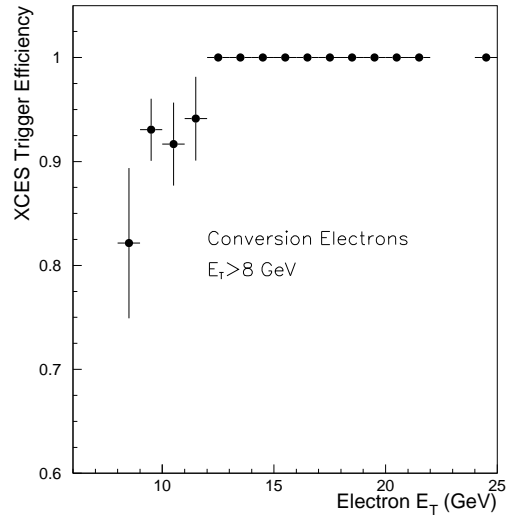}
 \label{xces_eff}
 \caption{Efficiency of the 8~GeV XCES electron trigger as a function
          of electron $E_T$, as measured in a conversion sample.}
\end{minipage}
\end{figure}

In conclusion, we have built and commissioned a shower maximum trigger
that has allowed us to reduce the Level-2 electron trigger cross section
by a factor of two while remaining highly efficient for good electrons.
Compared to the alternative of prescaling the old electron trigger, this
trigger will allow CDF to collect an additional 1-2~million electrons
from $b$ decays in the 100~pb$^{-1}$ of luminosity expected in 1994-95.

This work is supported in part by the U.S. Department of Energy,
Division of High Energy Physics, Contracts DE-AC02-76ER01112 and
W-31-109-ENG-38.


\begin{thebibliography}{99}

\bibitem{lum} As this article was being prepared, a world-record $\bar{p}p$
              luminosity of 1.28$\times10^{31}$cm$^{-2}$sec$^{-1}$
              was achieved at the Tevatron on July 23, 1994.

\bibitem{cdf_det} F. Abe {\em et al.}, Nucl. Instrum. Methods Phys. Res.
                  {\bf A271}, 387 (1988) and references therein.

\bibitem{cdf_trig} D. Amidei {\em et al.}, Nucl. Instrum. Methods Phys. Res.
                   {\bf A269}, 51 (1988).

\bibitem{rabbit} The CDF front-end system is described in
                 G. Drake {\em et al.}, Nucl. Instrum. Methods Phys. Res.
               {\bf A269}, 68 (1988).
\end{thebibliography}
\end{document}